\documentclass[bibnotes,twocolumn, showpacs,aps,prb,superscriptaddress,a4paper]{revtex4}

\usepackage{color,epsfig,rotating,graphicx,subfigure}
\usepackage{amsmath,amssymb,amscd}

\usepackage[latin1]{inputenc}
\usepackage[english]{babel}

\usepackage{dsfont}
\usepackage{textcomp}

\renewcommand{\Re}{{\rm Re}}
\renewcommand{\Im}{{\rm Im}}

\newcommand{\ri}{{\rm i}}
\newcommand{\re}{{\rm e}}
\newcommand{\rd}{{\rm d}}
\newcommand{\rr}{{\rm r}}
\newcommand{\rT}{{\rm T}}
\newcommand{\rL}{{\rm L}}  
\newcommand{\rh}{{\rm h}}

\newcommand{\rs}{{\rm s}}
\newcommand{\rp}{{\rm p}}  

\newcommand{\sgn}{{\rm sgn}} 

\newcommand{\kb}{k_{\rm B}}

\newcommand{\Tr}{{\rm Tr}}

\newcommand{\rth}{{\rm th}}

\begin{document}

%
%
\title{Nanoscale heat flux between anisotropic uniaxial media}

\author{S.-A. Biehs} 
\affiliation{Laboratoire Charles Fabry, Institut d'Optique, CNRS, Universit\'{e} Paris-Sud, Campus
Polytechnique, RD128, 91127 Palaiseau Cedex, France}

\author{P. Ben-Abdallah}
\affiliation{Laboratoire Charles Fabry, Institut d'Optique, CNRS, Universit\'{e} Paris-Sud, Campus
Polytechnique, RD128, 91127 Palaiseau Cedex, France}

\author{F. S. S. Rosa}
\affiliation{Laboratoire Charles Fabry, Institut d'Optique, CNRS, Universit\'{e} Paris-Sud, Campus
Polytechnique, RD128, 91127 Palaiseau Cedex, France}

\author{K. Joulain} 
\affiliation{Institut P', CNRS-Universit\'e de Poitiers UPR 3346, 86022 Poitiers Cedex, France}

\author{J.-J. Greffet}
\affiliation{Laboratoire Charles Fabry, Institut d'Optique, CNRS, Universit\'{e} Paris-Sud, Campus
Polytechnique, RD128, 91127 Palaiseau Cedex, France}

\date{21.02.2011}
\begin{abstract}
We present a theoretical study of near-field heat transfer between two 
anisotropic materials separated by a small vacuum gap and maintained in a
stationary non-equilibrium thermal situation. By combining standard stochastic 
electrodynamics and the Maxwell-Garnett description for effective media, we 
show that heat flux can be significantly enhanced by air inclusions.
This result is explained by : (a) the 
presence of extraordinary surface waves that give rise to supplementary 
channels for heat transfer throughout the gap, (b) an increase in the contribution
given by the ordinary surface waves at resonance, (c) and the appearance of 
frustrated modes over a broad spectral range.
\end{abstract}

\maketitle

\newpage

%
%

\section{Introduction}

Near field heat transfer~\cite{PvH1971,JoulainEtAl2005,VolokitinPersson2007,VinogradovDorofeyev2009} between closely spaced isotropic media has 
been intensively studied since it has been predicted that 
the heat flux at nanoscale can exceed the far-field limit of the 
Planck's blackbody theory by orders of magnitude~\cite{LevinEtAl1980,LoomisMaris1994}. When considering dielectrics, 
surface phonon polaritons provide additional enhancement as discussed in Refs.~\cite{MuletEtAl2001,Zhang2005, BiehsEtAl2010}.
Several experiments have recently confirmed the theoretical predictions for simple systems~\cite{ShenEtAl2008,RousseauEtAl2009,Kittel,HuEtAl2008,NarayaEtAl2008}.

With the modern techniques of nanofabrication it is now possible to 
explore a whole new level of complexity in material science and to 
fabricate artificial materials that can exhibit a considerable diversity 
of optical properties~\cite{Pendry2000,Pendry2004,FangEtAl2005,Shalaev2007,JacobEtAl2010,FengEtAl2010}. 
In many situations, these composite media possess 
privileged orientations so that their electromagnetic response depends on 
the direction of photons propagation. When the photon's wavelength
in such a medium is large compared to the size of its 
representative unit cell, the latter behaves effectively like an anisotropic 
material and therefore may be described by an effective 
permittivity tensor (and, when necessary, an effective permeability as well). This naturally 
points to the question of how anisotropy influences the near-field heat transfer. 

In this work, we address this question in the particular case of two 
semi-infinite uniaxial media characterized by optical axes orthogonally
 oriented with respect to the surface 
of interaction. The paper is organized as follows: In Sec.~II we derive
the expression for the heat flux between two anisotropic media.
After a brief description of the relevant composite media to our purposes in Sec.~III 
we investigate in Sec.~IV the surface and Brewster modes supported by them and their main features. 
Next, we compare in Sec.~V the near-field heat exchanges between two 
uniaxial media to the classical ones between two isotropic media. 
Finally, in order to explain the difference in the behavior of isotropic and anisotropic materials, we discuss in Sec.~VI the transmission factor in detail between two uniaxial media and in Sec.~VII
we present our conclusions.

%
%

\section{Radiative heat transfer between anisotropic media}

Let $B_1$ and $B_2$ be two anisotropic semi-infinite bodies, filling respectively 
the regions $z < 0$ and $z > d$ and leaving a vacuum gap between them.
In order to ensure a stationary process, we assume that
the $B_i$ are in local thermal equilibrium at a temperature $T_i$, with $T_1 \neq T_2$. 
The heat flux ${\rm P}$ between the two bodies is given by
\begin{equation}
  {\rm P}(T_1,T_2,a) = \int_{A_{12}} \!\! {\bf dA} \cdot \langle {\bf S} \rangle ,
\end{equation}
where ${\bf S} = {\bf E} \times {\bf H}$ is the Poynting
vector and $A_{12}$ is any surface that separates the two bodies.
By taking such a surface to be a plane defined by $z=z_0 \, (0<z_0<d)$
and using the (transverse) translational invariance of our system,
the previous equation simplifies to
\begin{equation}
  {\rm P}(T_1,T_2,a) = A \langle S_z \rangle ,
\end{equation}
showing that only the $z$-component of the Poynting vector is needed.
After a straightforward calculation, the latter can be conveniently written as~\cite{VolokitinPersson2007}
\begin{equation}
  \langle S_z \rangle =  \int_0^\infty\!\frac{\rd \omega}{2 \pi} \bigl[ \Theta(\omega,T_1) - \Theta(\omega,T_2)\bigr] \langle S_\omega \rangle ,
\label{Eq:MeanPoynting}
\end{equation}
where we identify the mean energy of a harmonic oscillator
\begin{equation}
  \Theta(\omega,T) = \frac{\hbar \omega}{\re^{\frac{\hbar \omega}{\kb T}} - 1} ,
\end{equation}
and also the averaged spectral Poynting vector~\cite{VolokitinPersson2007}
\begin{equation}
\begin{split}
  \langle S_\omega \rangle &= 2\, \Re\, \Tr\, \biggl[ \int_{A}\!\!\rd {\bf r}_{\|}' \biggl( \mathds{G}(\mathbf{r,r'}) \partial_z \partial_z'\mathds{G}^\dagger(\mathbf{r,r'})  \\
                          &\qquad\qquad\qquad - \partial_z \mathds{G}^\dagger(\mathbf{r,r'}) \partial_z' \mathds{G}(\mathbf{r,r'}) \biggr) \biggr]_{z' = z = z_0}.
\end{split}
\end{equation}
where ${\bf r} = {\bf r}_{\|} + z\hat{z}$ and $\mathds{G}(\mathbf{r,r'})$ is the electrical Green's dyadic, satisfying 
\begin{eqnarray}
\label{GreenEquation}
&&\left[ \overrightarrow{\nabla} \!\times\! \overrightarrow{\nabla} \!\times - \frac{\omega^2}{c^2} \epsilon({\bf r},\omega) \right] {\mathds G}({\bf r},{\bf r}',\omega) = \delta({\bf r}-{\bf r}') \mathbb{I} .
\end{eqnarray}
Moreover, we have introduced Boltzmann's constant $\kb$, Planck's constant $2 \pi \hbar$; the 
$\dagger$ symbolizes hermitian conjugation and $\Tr$ the $3 \times 3$ trace. 

In order to evaluate the heat flux in the given geometry we have to determine the Green's dyadic inside the gap region. This can be done by considering the multiple scattering of a plane wave due to a source inside the gap~\cite{Philbin08}. Details and the final
expression for the Green's dyadic can be found in appendix~\ref{App:Greensfunction}. When inserting the resulting expression in Eq.~(\ref{Eq:IntracavityGreen})
into the heat flux formula, we find
\begin{equation}
  \langle S_\omega \rangle = \int\!\!\frac{\rd^2\kappa}{(2 \pi)^2} \, T(\omega,\boldsymbol{\kappa}; d).
\label{Eq:SpectralPoynting}
\end{equation}
The integral is carried out over all transverse wave vectors 
$\boldsymbol{\kappa} = (k_x,k_y)^{\rm t}$ including 
propagating modes with $\kappa < \omega/c$ and evanescent modes with $\kappa > \omega/c$, where $c$ is the velocity of light in vacuum. The energy transmission coefficient $T(\omega,\boldsymbol{\kappa}; d)$ is different for propagating and evanescent modes and can be stated as
\begin{equation}
\begin{split}
   T&(\omega,\boldsymbol{\kappa}; d) =  \\  
    &\begin{cases}
     \Tr\bigl[(\mathds{1} - \mathds{R}_2^\dagger \mathds{R}_2)  \mathds{D}^{12}(\mathds{1} - \mathds{R}_1^\dagger \mathds{R}_1)  {\mathds{D}^{12}}^\dagger \bigr], & \kappa < \omega/c\\
     \Tr\bigl[(\mathds{R}_2^\dagger - \mathds{R}_2) \mathds{D}^{12} (\mathds{R}_1 - \mathds{R}_1^\dagger)  {\mathds{D}^{12}}^\dagger \bigr]\re^{-2 |\gamma_{\rr}| d},  & \kappa > \omega/c
  \end{cases}
\end{split} 
\label{Eq:TransmissionCoeff}
\end{equation}
where $\gamma_{\rr} = \sqrt{\omega^2/c^2 - \kappa^2}$ and $\mathds{R}_1$, 
$\mathds{R}_2$ are the $2 \times 2$ reflection matrices characterizing interfaces.
By writing them a bit more explicitly, 
\begin{equation}
\label{ReflectionMatrices}
{\mathds R}_i = \left[
\begin{array}{cc}
   r^{{\rm s,s}}_i (\omega, \kappa) &  r^{{\rm s,p}}_i (\omega, \kappa) \\
   r^{{\rm p,s}}_i (\omega, \kappa) &  r^{{\rm p,p}}_i (\omega, \kappa) 
\end{array} \right] ,
\end{equation}
we see that their elements $r_i^{\lambda,\lambda'}$ are the 
reflection coefficients for the scattering of an incoming $\lambda$-polarized plane wave  
into an outgoing $\lambda'$-polarized wave. In the isotropic limit they reduce to the 
usual Fresnel coefficients
\begin{eqnarray}
\label{FresnelCoefficients}
&&r^{{\rm s,s}}_i(\omega, \kappa) = \frac{\gamma_{\rr} - \sqrt{\epsilon_i(\omega) \omega^2/c^2- \kappa^2}}{\gamma_{\rr}
 + \sqrt{\epsilon_i(\omega) \omega^2/c^2 - \kappa^2}} , \nonumber \\
&&r^{{\rm p,p}}_i(\omega, \kappa) = \frac{\epsilon_i(\omega) \gamma_{\rr} - \sqrt{\epsilon_i(\omega) \omega^2/c^2 - \kappa^2}}{\epsilon_i(\omega) \gamma_{\rr} + \sqrt{\epsilon_i(\omega) \omega^2/c^2-\kappa^2}} , \nonumber \\
&&r^{{\rm s,p}}_i(\omega, \kappa) = r^{{\rm p,s}}_i(\omega, \kappa) = 0 ,
\end{eqnarray}
and we see that the matrices become diagonal. In addition, we have introduced 
the matrix $\mathds{D}^{12}$, defined by
\begin{equation}
  \mathds{D}^{12} = (\mathds{1} - \mathds{R}_1 \mathds{R}_2 \re^{2 \ri \gamma_{\rr} d})^{-1},
\end{equation}
which gives rise to a Fabry-P\'{e}rot-like denominator for $T(\omega,\boldsymbol{\kappa}; d)$ in the isotropic case.

From Eqs~(\ref{Eq:MeanPoynting}), (\ref{Eq:SpectralPoynting}) and 
(\ref{Eq:TransmissionCoeff}) we see that, once the reflection matrices are known,
it is possible to determine the heat flux between two arbitrary 
anisotropic semi-infinite bodies kept at fixed temperatures $T_1$ and $T_2$. 
Moreover, in order to have an independent check, we verified that Eq.~ (\ref{Eq:SpectralPoynting}) also can
be derived from the general scattering formalism derived on Ref.~\cite{Bimonte2009}. 
In the following we will use these expressions to discuss the heat flux between two 
uniaxial anisotropic materials with their optical axes normal to the interface.

\begin{figure}[Hhbt]
\epsfig{file=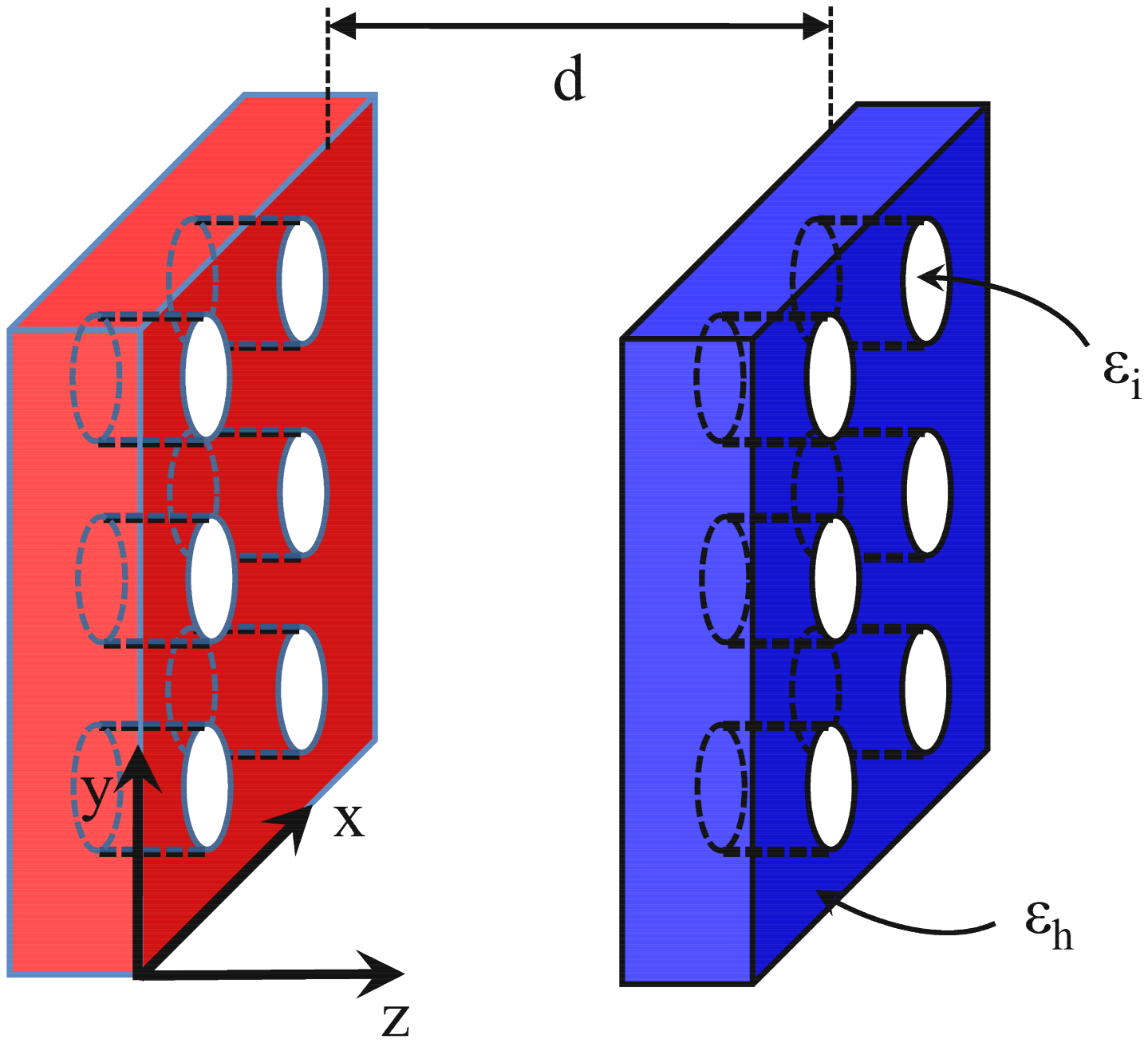, width=0.45\textwidth}
  \caption{\label{Fig:slabs} Sketch of two porous slabs with different temperatures separated by a vacuum gap. } 
\end{figure}

%
%

\section{Porous Materials}

The structures investigated in this paper are depicted in Fig.~\ref{Fig:slabs}. 
They are two semi-infinite media composed by 
a host isotropic material, defined by its complex dielectric function 
$\epsilon_\rh(\omega) = \epsilon_\rh'(\omega) + \ri \epsilon_\rh''(\omega)$ (where $\epsilon_\rh''(\omega) > 0$), with uniform
cylindrical inclusions oriented in the direction orthogonal to the surface as shown in Fig.~\ref{Fig:slabs}. 
These inclusions in turn are filled by a medium of dielectric permittivity $\epsilon_\ri$, that is also assumed to be isotropic. 
When the size of the representative unit cell is much smaller than all the other characteristic scales involved, a suitable 
volume average of the material's local electromagnetic response can be made. In our case, the emerging azimuthal symmetry in this 
long wavelength limit gives rise to effective uniaxial crystals with a permittivity tensor of the form
\begin{equation}
  \boldsymbol{\epsilon} = \epsilon_\parallel \left[ \mathbf{e}_x \otimes \mathbf{e}_x +  \mathbf{e}_y \otimes \mathbf{e}_y \right]
             +  \epsilon_\perp \mathbf{e}_z \otimes \mathbf{e}_z   
\end{equation}
where $\mathbf{e}_x$, $\mathbf{e}_y$, and $\mathbf{e}_z$ are orthogonal unit vectors in $x$, $y$, and $z$ direction. The parallel and perpendicular components can be derived from the Maxwell-Garnett effective medium theory (EMT)~\cite{SaarinenEtAl2008,ElserEtAl2006}
\begin{align}
  \label{Eq:eps_par}
  \epsilon_\parallel &= \epsilon_\rh \frac{\epsilon_\ri (1 + f) + \epsilon_\rh (1 - f)}{\epsilon_\ri (1 - f) + \epsilon_\rh (1 + f)}, \\
  \label{Eq:eps_perp}
  \epsilon_\perp &= \epsilon_\rh (1 - f) + \epsilon_\ri f,
\end{align}
where $f$ is the volume fraction of inclusions. For the structure considered in this work the deviation from the exact result of homogenization given in Refs.~\cite{HaleviEtAl1999,KrokhinEtAl2002} is small even for relatively high filling factors such as $f = 0.5$.
Hence, we will discuss the heat flux between porous media with the Maxwell-Garnett expression for $f \in [0,0.5]$ in this work.

The condition of long wavelengths sets a limit to the lattice constant $a$ of the inclusions for which the EMT can be used. In the
far-field regime this condition is fullfilled when the thermal wavelength $\lambda_\rth = \hbar c /\kb T$ is larger than $a$. 
In the near-field region the contributing modes at a distance $d$ above the porous material have a lateral wavelength which depends on $d$.  
For $\kappa = 2\pi/a$ (which corresponds to a lateral wavelength $a$) the evanescent waves are damped 
as $\exp[ - \sqrt{(2 \pi /a)^2 - \omega^2/c^2} d] \approx \exp( -(2 \pi / a) d) $ in the non-retarded near-field region above the porous material. 
It follows that the contribution to the heat flux is dominated by evanescent waves with lateral wavelength larger than $a$ if $d > a /(2 \pi)$. 
On the other hand, one can argue that a nonlocal model for the permittivity is necessary if the lateral wave vectors $\kappa$ are on the order of $\pi/a$.
Since the exponential in the transmission coefficient in Eq.~(\ref{Eq:TransmissionCoeff}) for $\kappa > \omega/c$ sets a cutoff for $\kappa$ of the modes contributing 
to the near-field heat flux which is $\approx 1/d$, one finds that a local EMT description is permissible if $d > a / \pi$.
Hence, for a given lattice constant $a$ of the inclusions the validity of the EMT in Eq.~(\ref{Eq:eps_par}) in the near 
field regime is restricted to $d > a / \pi$. Artificial structures as depicted in Fig.~\ref{Fig:slabs} can have an $a$ on the order of $100\,{\rm nm}$~\cite{FengEtAl2010}
so that the distances for which the EMT can be considered as appropriate in this case are about $d > 30\,{\rm nm}$. 
Nonetheless, chemically produced nanoporous materials can show smaller structures~\cite{ChuangEtAl1989} so that we will consider distances $d \in [10\,{\rm nm},100\,\mu {\rm m}]$.


\begin{figure}[Hhbt]
  \epsfig{file=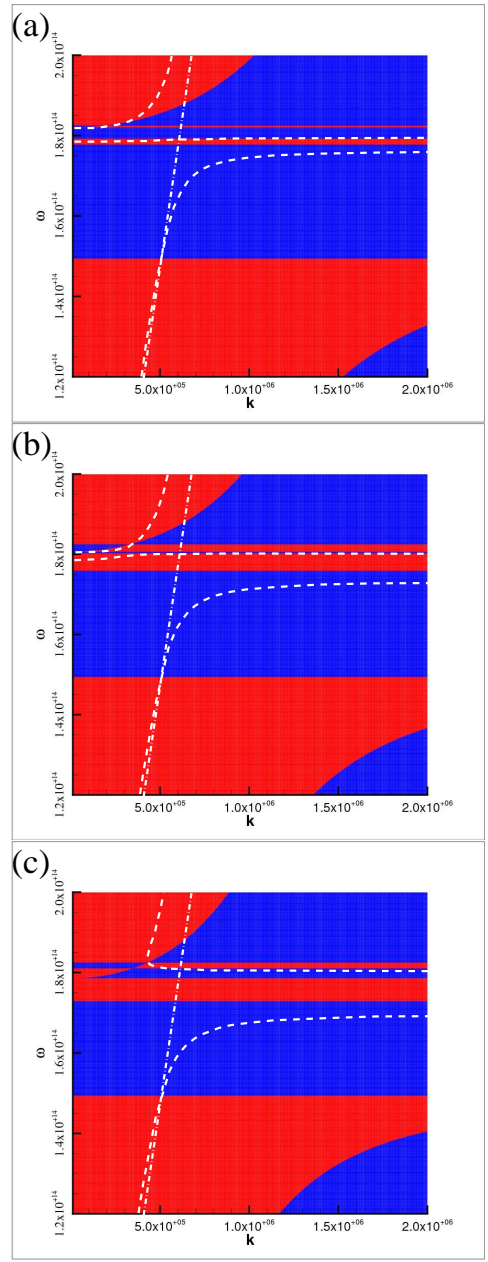, width=0.35\textwidth}
  \caption{\label{Fig:disp} Plot of the dispersion curves (white dashed lines) from Eq.~(\ref{Eq:Disp}) in 
                the ($\omega$,$\kappa$) plane for filling factors (a) $f = 0.1$, (b) $f = 0.3$, and (c) $f = 0.5$.
                The white dash-dotted line represents the light line in vacuum ($\omega = \kappa c$). 
                Furthermore the dark (blue) areas mark the region for which $\gamma_\rp$ is purely real, whereas
                the bright (red) areas are the regions for which $\gamma_\rp$ is purely imaginary.  } 
\end{figure}

\begin{figure}[Hhbt]
  \epsfig{file=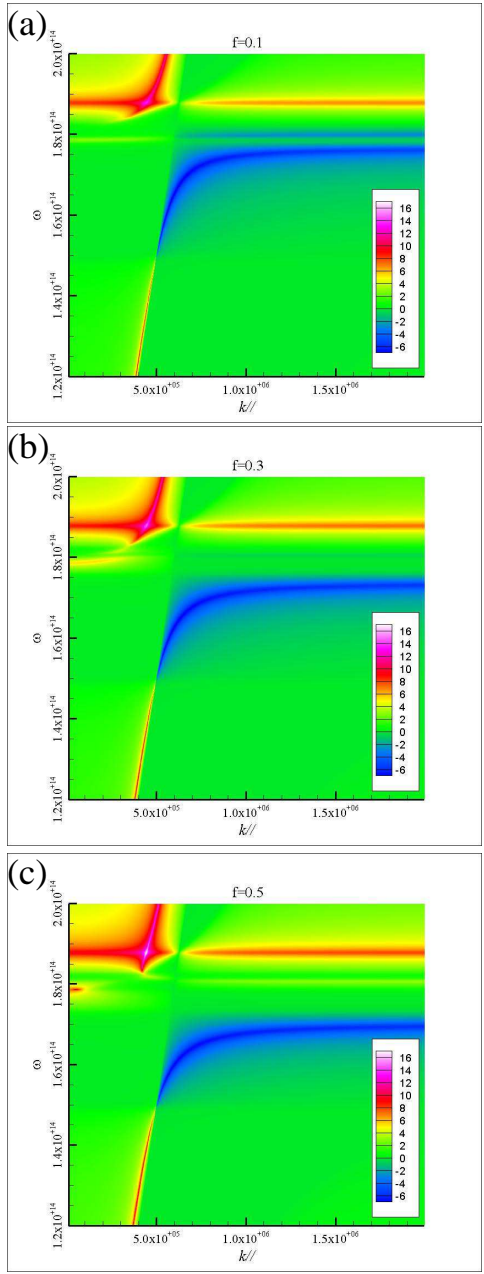, width=0.35\textwidth}
  \caption{\label{Fig:rfilling} Plot of $\ln(1/|r_{\rp,\rp}|^2)$ in $(\omega,\kappa)$ plane for (a) $f = 0.1$, (b) $f = 0.3$, and (c) $f = 0.5$.
          } 
\end{figure}


\section{Surface and Brewster modes in porous Media}

Let us study the surface waves supported by these media when they are sufficiently far away from each other so that any coupling of evanescent waves can be neglected. By definition, these surface waves are resonant surface 
modes and therefore are determined by the poles of the reflection coefficients of these media. For out-of-plane uniaxial media the components of the reflection matrix are 
\begin{eqnarray}
 && r_{\rs, \rs} (\omega, \kappa)= \frac{\gamma_{\rr} - \gamma_\rs}{\gamma_{\rr} - \gamma_\rs} ,\\
 && r_{\rp, \rp} (\omega, \kappa)= \frac{\epsilon_\parallel \gamma_{\rr} - \gamma_\rp}{\epsilon_\parallel \gamma_{\rr} + \gamma_\rp} ,\\
  &&r_{\rs, \rp} = r_{\rp, \rs} = 0 ,
\label{Eq:Rp}
\end{eqnarray}
where $\gamma_{\rs,\rp}$ are given by the solutions of Fresnel equations
in the anisotropic material~\cite{Yeh1988}
\begin{eqnarray}
&&\gamma_\rs = \sqrt{\epsilon_{\|}\omega^2/c^2 - \kappa^2} ,\\
&&\gamma_\rp = \sqrt{\epsilon_{\|}\omega^2/c^2 - \frac{\epsilon_{\|}}{\epsilon_{\bot}}\kappa^2} ,
\end{eqnarray}
and hence it follows at once that the surface modes are determined by
\begin{eqnarray}
\label{surfacemodes}
&&(\gamma_{\rr} + \gamma_\rs) = 0 ,\\
&&(\epsilon_\parallel \gamma_{\rr} + \gamma_\rp) = 0.
\end{eqnarray}
It is straightforward to verify that in this case only the second equation above 
can be satisfied, meaning that only p-polarized surface waves can exist at the
interface of these media. Solving that equation explicitly for $\kappa$ gives us the sought  dispersion relation of surface waves 
\begin{equation}
  \kappa = \frac{\omega}{c} \sqrt{\frac{\epsilon_\perp (\epsilon_\parallel - 1)}{\epsilon_\parallel \epsilon_\perp - 1}}.
\label{Eq:Disp}
\end{equation}
but one must be aware that (\ref{Eq:Disp}) has {\it two} branches, and only one is connected
to surface modes\footnote{The other branch is connected with the so called Brewster 
modes \cite{Archambault09}, that are propagating waves for which $r_{\rp,\rp}$ vanishes.}. Since their dispersion relation involves $\epsilon_\parallel$ and $\epsilon_\perp$, these waves are also called extraordinary surface waves~\cite{Liscidini10}, and they reflect 
the material anisotropy. When $\epsilon_\parallel = \epsilon_\perp = \epsilon$, Eq.~(\ref{Eq:Disp}) degenerates into the well-known dispersion 
relation $\kappa = \omega/c \sqrt{\epsilon/(\epsilon + 1)}$ of surface modes supported by a semi-infinite isotropic medium (bounded by vacuum) with
 a dielectric permittivity $\epsilon$. In Fig.~\ref{Fig:disp} we plot the dispersion curves 
for silicon carbide (SiC) with vacuum inclusions for different 
filling factors $f = 0.1$, $f = 0.3$ and $f = 0.5$. The dielectric function of SiC is described~\cite{Shchegrov00} by the simple model
\begin{equation}
  \epsilon_\rh = \epsilon_\infty \frac{\omega^2 - \omega_\rL^2 - \ri \omega \Gamma}{\omega^2 - \omega_\rT^2 - \ri \omega \Gamma}
\end{equation}
where $\omega_\rL = 1.827\cdot10^{14}\,{\rm s}^{-1}$, $\omega_\rT = 1.495\cdot10^{14}\,{\rm s}^{-1}$, $\Gamma = 0.9\cdot10^{12}\,{\rm s}^{-1}$, and $\epsilon_\infty = 6.7$ 
denote respectively the longitudinal and transversal optical phonon pulsation, the damping factor and the high frequency dielectric constant, respectively. In order to avoid the inherent difficulties of multiple possible interpretations of complex dispersion relations~\cite{Archambault09}, we have deliberately neglected the host material losses to represent these curves. The relevance of this approximation can be checked by comparing Fig.~\ref{Fig:disp} with Fig.~\ref{Fig:rfilling}, where we plot the reflection coefficients of dissipating porous material. In order to distinguish between evanescent and 
propagative waves inside the effective medium, solutions of Eq.~(\ref{Eq:Disp}) are superimposed in Fig.~\ref{Fig:disp} to a two-color background. This background is a binary representation of $\zeta = \sgn(\epsilon_\parallel \omega^2/c^2 - \epsilon_\parallel \kappa^2 /\epsilon_\perp)$. In the blue zones $\zeta < 0$ so that only evanescent modes can exist, and conversely, in the red zones we have $\zeta > 0$ and all modes are propagative. Similarly the light line $\omega = c \kappa$ allows us to distinguish between the radiative (propagative) and the non-radiative (evanescent) modes inside the vacuum. Notice that, in order to satisfy Eq.~(\ref{Eq:Disp}) both $\zeta$ and $\sgn(\omega^2/c^2 - \kappa^2)$ must be the same. 
In other words, frustrated modes cannot satisfy the dispersion relation (\ref{Eq:Disp}).

Now let us turn to the description of modes supported by our artificial structures. For low filling factors we note in Fig.~\ref{Fig:disp} (a) the existence of two surface modes. The first one (at a lower frequency) corresponds to the classical surface phonon-polariton (SPP) supported by 
a massive SiC sample~\cite{Raether}. That surface mode is also present in isotropic SiC. The most interesting feature of  Fig.~\ref{Fig:disp} (a) is, however, the appearance of a second surface mode at higher frequencies, because it is a signature of the anisotropic character of the material and therefore a direct consequence of the vacuum inclusions in the host medium. As the porosity increases, both surface waves split. Beyond a critical 
filling factor between $f = 0.3$ and $f = 0.5$, the upper surface wave disappears as is seen in Fig.~\ref{Fig:rfilling}. 
Nevertheless the SPP which still exists continue to move 
toward the smaller frequencies, i.e., to $\omega_\rT$. 
Above the light line, we see that the anisotropy gives rise to two different types of Brewster modes. At high frequency we recognize the usual modes where the reflection coefficient [Fig.~\ref{Fig:rfilling} (a)] of the effective medium vanishes. 
In addition to these modes, different Brewster modes appear depending on the value of filling factor. Also, we see on the reflection curves (Fig.~\ref{Fig:rfilling}) that the Christiansen point \cite{BohrenHuffman} for which the reflectivity is zero for all $\kappa$ 
does not depend on the porosity. Indeed, an inspection of expressions 
(\ref{Eq:eps_par}) and (\ref{Eq:eps_perp}) shows that the condition for the Christiansen point of the host material $\epsilon_\rh = 1$ implies that $\epsilon_\parallel = 1$ and $\epsilon_\perp = 1$ so that, according to (\ref{Eq:Rp}), the reflection coefficients vanish.


\section{Heat flux between porous media}

\begin{figure}[Hhbt]
  \epsfig{file=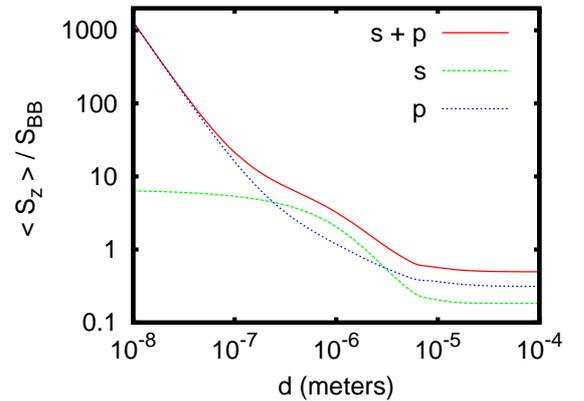, width=0.45\textwidth}
  \caption{\label{Fig:PoyntingIsotrop} Heat flux between two SiC plates over distance 
           with $T_1 = 300\,{\rm K}$ and $T_2 = 0\,{\rm K}$. The flux is normalized to
           the value for two black bodies $S_{\rm BB} = 459.6\,{\rm W}{\rm m}^{-2}$. 
           The contribution of the s- and p-polarized part are shown as well. 
          } 
\end{figure}

Before we discuss the influence of the inclusions on the heat flux, we show
in Fig.~\ref{Fig:PoyntingIsotrop} the results of the mean Poynting 
vector $\langle S_z \rangle$ between two semi-infinite SiC bodies at fixed temperatures
$T_1 = 300\,{\rm K}$ and $T_2 = 0\,{\rm K}$. First of all one can see that 
the heat flux becomes very large for distances much smaller than the thermal 
wavelength $\lambda_\rth = \hbar c / \kb T$ (which is about $7.68\, \mu{\rm m}$ for $T = 300\,{\rm K}$). At $d = 10\,{\rm nm}$ the heat flux for the two SiC bodies is about $1000$ times larger than the heat flux between two black bodies. This increase is due to the frustrated total internal reflection and to the coupled surface phonon polariton modes~\cite{BiehsEtAl2010}.
In the propagating regime, i.e., for distances larger than $\lambda_\rth$ the heat flux is determined by Kirchhoff-Planck's law and is limited by the black-body value. 
Note, that the heat flux is dominated by the p-polarized modes for distances smaller than
$100\,{\rm nm}$ and larger than $10\,\mu{\rm m}$, whereas
for distances in between it is dominated by the s-polarized modes.

\begin{figure}[Hhbt]
 \epsfig{file=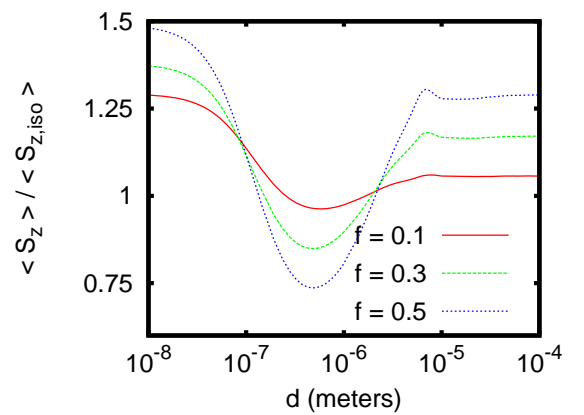, width=0.45\textwidth}
  \caption{\label{Fig:PoyntingPorous} Heat flux between two porous SiC plates over distance 
           with $T_1 = 300\,{\rm K}$ and $T_2 = 0\,{\rm K}$. The flux is normalized to
           the value for two SiC plates shown in Fig.~\ref{Fig:PoyntingIsotrop}. 
          } 
\end{figure}

Now, we introduce the inclusions by using the Maxwell-Garnett expression in Eq.~(\ref{Eq:eps_par}) and (\ref{Eq:eps_perp}). We use the same filling factor for both materials, so that
we have a symmetric situation. In Fig.~\ref{Fig:PoyntingPorous} we show the resulting heat flux
normalized to the values for the two non-pourous SiC plates shown in Fig.~\ref{Fig:PoyntingIsotrop}. We find that for distances smaller than $100\,{\rm nm}$ and larger than $1\, \mu{\rm m}$ the heat flux becomes larger when we add air inclusions, whereas for intermediate distances the heat flux is reduced. 

\begin{figure}[Hhbt]
  \epsfig{file=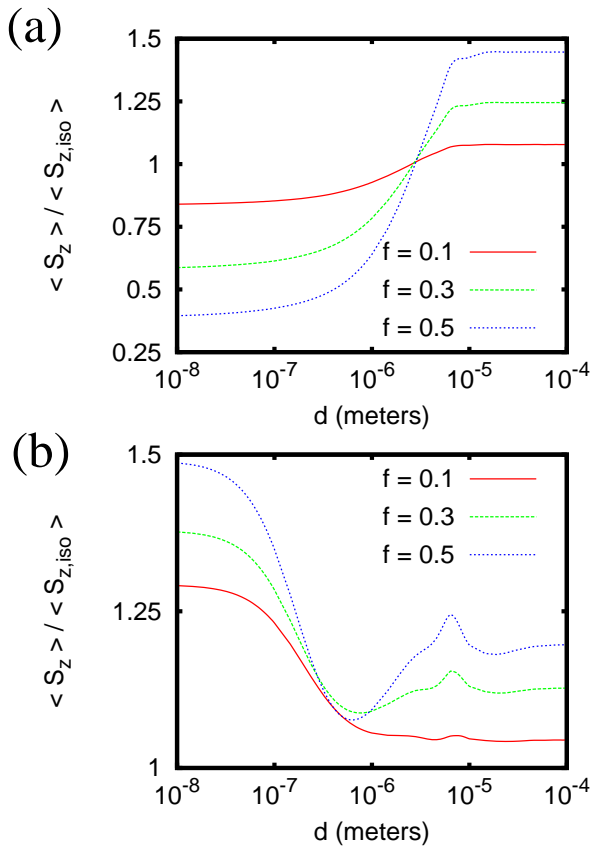, width=0.45\textwidth}
  \caption{\label{Fig:PoyntingPoroussp} 
          As in Fig~\ref{Fig:PoyntingPorous} but for the (a) s- and (b) p-polarized contribution only.
          } 
\end{figure}

In order to see how the s- and p-mode
contribution is changed by the porosity, we show in Fig.~\ref{Fig:PoyntingPoroussp}
(a) and (b) the plots for the separate contributions of s- and p-polarized modes. It is clear that the p-polarized part of the heat flux gets enhanced for all distances when compared to the isotropic case, regardless of the filling factor. The s-polarized part in turn gives a larger heat flux for distances larger than about $1\, \mu{\rm m}$ and a smaller heat flux for distances smaller than $1\, \mu{\rm m}$. Therefore, the smaller heat flux found in Fig.~\ref{Fig:PoyntingPorous} for intermediate distances is associated to the dominance of s-polarized modes in that distance regime. 

In summary, by introducing inclusions we find for large and small
distances an increase of the heat flux. For the propagating regime ($d > \lambda_\rth$) this can be understood from a simple argument: the vacuum holes simply dilute the material so that, according to Kirchhoff's law,  the reflectivity is decreased and hence the emissivity is increased. In fact, for $f = 1$ one would retrieve the black body result, since in this case the reflectivity is zero. On the other hand, there is no such simple argument for the increased heat flux in the near-field region. Here, it is necessary to study how the coupled surface modes, which give the main contribution to the heat flux for distances smaller than $100\,{\rm nm}$, are influenced by the introduction of the inclusions. This will be done by inspection of the transmission coefficient in the next section.


\section{Transmission coefficient}

As mentioned before, for the small distance regime ($d < 100\,{\rm nm}$) the heat flux between two isotropic semi-infinite SiC-bodies is solely dominated by the p-polarized contribution. This remains true for the porous SiC bodies. In fact, the dominance of the p-polarized contribution becomes even greater with increasing filling factors. Hence, to understand the observation that by introducing some porosity the heat flux becomes larger, it suffices to study the p-polarized contribution. 

\begin{figure*}[Hhbt]
  \epsfig{file=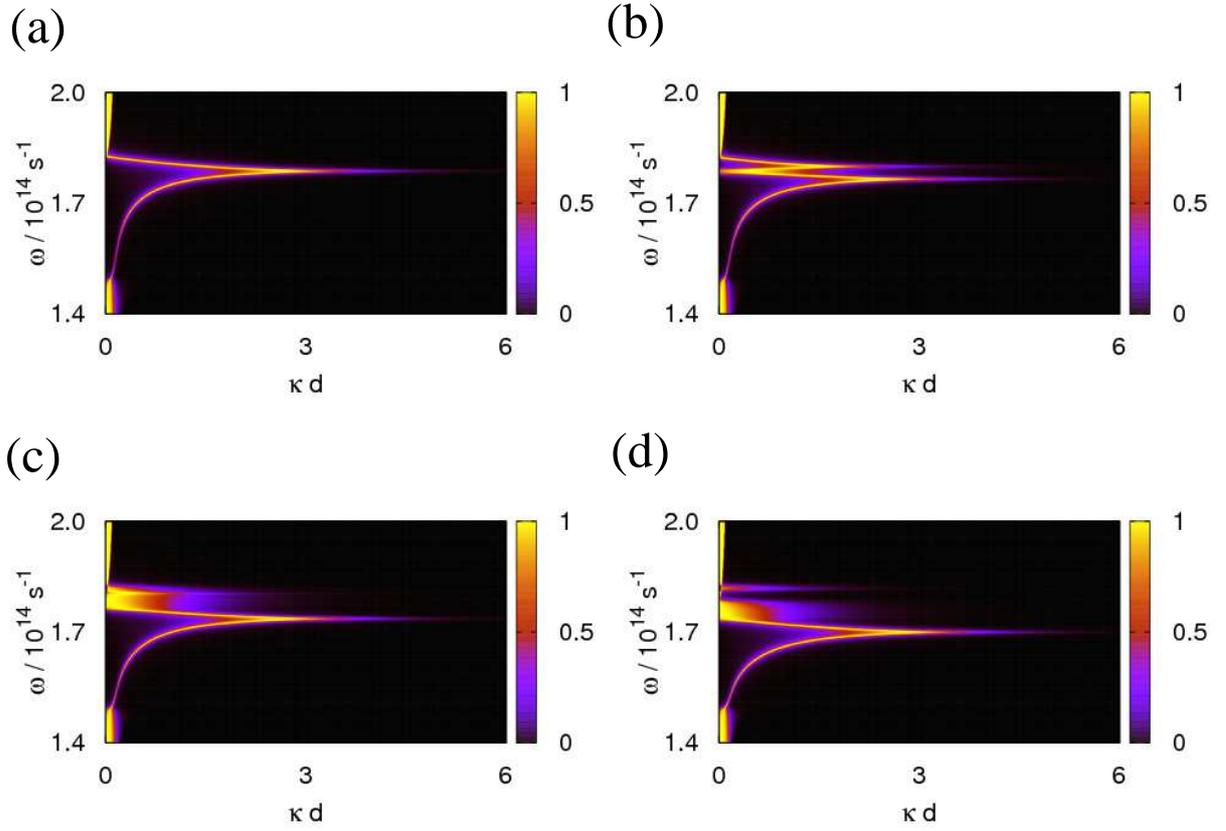, width=0.9\textwidth}
  \caption{\label{Fig:Transmission_omega_kappa} Transmission coefficient $T_\rp(\omega,\kappa;d)$ in the $(\omega,\kappa)$-plane 
           for two porous SiC slabs with different filling factors (a) $f = 0$, (b) $f = 0.1$, (c) $f = 0.3$, and (d) $f = 0.5$.
           The distance is fixed at $d = 100\,{\rm nm}$.}
\end{figure*}

In Fig.~\ref{Fig:Transmission_omega_kappa} we show the transmission coefficient $T_\rp(\omega,\kappa;d)$ in the $(\omega,\kappa)$-plane 
for different filling factors and a distance $d = 100\,{\rm nm}$. In Fig.~\ref{Fig:Transmission_omega_kappa} (a) one can 
see $T_\rp(\omega,\kappa;d)$ for two isotropic SiC plates. Here, $T_\rp(\omega,\kappa;d)$ is one or close to one
for the propagating modes, the total internal reflection modes and the coupled surface phonon polaritons. 
In the plotted region one can mainly see the coupled surface phonon polaritons, which are responsible for the large heat flux at 
small distances. Now, for $f = 0.1$ one can see in Fig.~\ref{Fig:Transmission_omega_kappa} (b) that a second coupled surface mode appears due to the air inclusions. In addition, the coupled surface mode of the bulk SiC is shifted to smaller frequencies. When increasing the filling factor [Fig.~\ref{Fig:Transmission_omega_kappa} (c) and (d)] the upper coupled surface modes shift to higher frequencies and become less important for the transmission coefficient. On the other hand, the low frequency surface modes shift further to lower frequencies. Between the two coupled surface mode branches a band of frustrated internal reflection modes is formed which gives also a non-negligible contribution to the transmission coefficient.

\begin{figure}[Hhbt]
  \epsfig{file=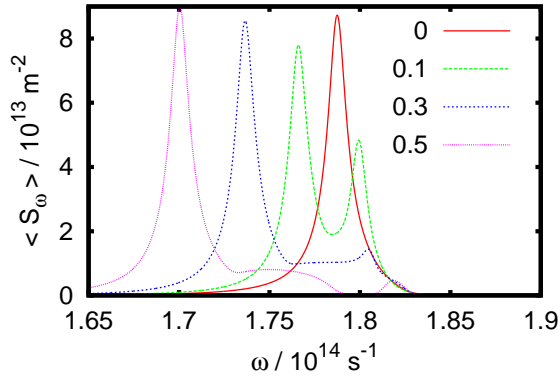, width=0.45\textwidth}
  \caption{\label{Fig:Sspectral} Spectral mean Poynting vector $\langle S_\omega \rangle$ defined in Eq.~(\ref{Eq:SpectralPoynting}) 
           for two porous SiC slabs with different filling factors $f = 0, 0.1 , 0.3, 0.5$ considering only the p-polarized contribution.
           The distance is fixed at $d = 100\,{\rm nm}$.}
\end{figure}

In order to get further information we now consider the spectral mean Poynting vector $\langle S_\omega \rangle$ defined in Eq.~(\ref{Eq:SpectralPoynting})
for p-polarization only. 
We have plotted this quantity in Fig.~\ref{Fig:Sspectral} at the same distance as before, i.e., $d = 100\,{\rm nm}$, and again
for different filling factors. As in Fig.~\ref{Fig:Transmission_omega_kappa} one can see the strong contribution of the two coupled surface mode resonances, which are shifted in frequencies when changing the filling factor. Furthermore, one can now observe, that when increasing the filling factor the low frequency resonance is not only shifted to smaller frequencies, but the resonance is also getting stronger. 

The study can now be completed when considering the mean transmission factor for the p-polarized modes, that was introduced in Ref.~\cite{BiehsEtAl2010} as
\begin{equation}
  \overline{T}_\rp(\kappa) = \frac{3}{\pi^2} \int_0^\infty \!\! \rd u\, f(u) T_\rp (u,\kappa;d)
\label{Eq:MeanT}
\end{equation}
with $u = \hbar \omega/\kb T$ and $f = u^2 \re^u/(e^u - 1)^2$. It represents the mean transmission coefficient of a mode specified by it's wave vector $\kappa$ for a given temperature $T$ and a small temperature difference $\Delta T$ between
the two bodies. By means of this quantity the heat flux can be rewritten in a Landauer-like form~\cite{BiehsEtAl2010} 
\begin{equation}
  \langle S_z \rangle = \frac{\pi^2}{3} \frac{\kb^2 T}{h} \int \!\! \frac{\rd \kappa}{2\pi}\, \kappa \, \overline{T}_\rp(\kappa) \Delta T.
\end{equation}
Note, that for $\kappa d \gg 1$ and $\kappa > \omega/c$ the transmission coefficient $T_\rp(\omega,\kappa;d)$ is exponentially 
damped [see Eq.~(\ref{Eq:TransmissionCoeff})] and therefore also the mean transmission factor $\overline{T}_\rp(\kappa)$. This
damping determines the wave vector cutoff and hence the number of states contributing to the heat flux.  

\begin{figure}[Hhbt]
  \epsfig{file=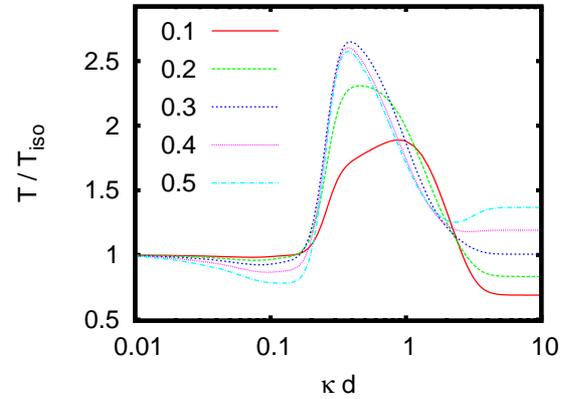, width=0.45\textwidth}
  \caption{\label{Fig:Mean} Mean transmission coefficient defined in Eq.~(\ref{Eq:MeanT}) for different filling factors 
           normalized to the isotropic case ($f = 0$).
           The distance is fixed at $d = 100\,{\rm nm}$ and the temperature at $T = 300\,{\rm K}$.}
\end{figure}

Now, in Fig.~\ref{Fig:Mean} we plot $\overline{T}_\rp(\kappa)$ for a given distance of $d = 100\,{\rm nm}$ and different 
filling factors normalized to the mean transmission factor for two semi-infinite SiC bodies. For $f = 0.1$ the
mean transmission coefficient for the porous SiC increases for intermediate $\kappa$ but decreases for very large $\kappa$.
The increased mean transmission factor is due to the second coupled surface mode and the frustrated modes, whereas the lower value for large wave vectors can be attributed to a stronger cutoff in the transmission coefficient, which means that the number of contributing modes is decreased.
The enhancement of the transmission factor due to the surface mode prevails and leads to an enhanced heat flux at that distance. The same
mechanism is responsible for the enhanced heat flux for $f = 0.3$. On the other hand, for larger filling factors the curves change slightly for intermediate $\kappa$ compared to the curve for $f = 0.3$. The contribution in that intermediate region is due to the second coupled surface mode branch and the frustrated modes. But for very large $\kappa$
the mean transmission coefficient increases compared to  $f = 0.3$. This means that by introducing a higher porosity we soften the cutoff of the transmission coefficient. Hence, the number of modes contributing to the heat flux is increased and results for large filling factors in a further enhanced heat flux.  

The dependence of the cutoff on the filling factor for large $\kappa$ can easily be discussed for the transmission coefficient $T_\rp(\kappa,\omega;d)$. It was found in Ref.~\cite{BiehsEtAl2010} that the cutoff region, i.e., where $T_\rp(\kappa,\omega;d)$ is exponentially damped, is given by
\begin{equation}
  \kappa_{\rm iso} > \log\biggl( \frac{2}{\Im(\epsilon)} \biggr) \frac{1}{2 d}
\end{equation}
when considering two isotropic semi-infinite bodies at the surface mode resonance frequency [see also Refs.~\cite{Joulain2010}].
For the uniaxial anisotropic case as considered here, this relation changes to
\begin{equation}
  \kappa_{\rm uni} > \log\biggl( \frac{2}{\Im(\sqrt{\epsilon_\parallel \epsilon_\perp})} \biggr) \frac{1}{2 d}
\end{equation}
where the permittivities have to be evaluated at the surface mode resonance frequency of the 
semi-infinite anisotropic body (see Appendix B). In Fig.~\ref{Fig:Cutoff} we show a plot of $\kappa_{\rm uni} / \kappa_{\rm iso}$
over the filling factor. It is seen that by introducing the air inclusions the cutoff first decreases and then monotonically increases. This is the same qualitative behavior as observed for the mean transmission factor $\overline{T}_\rp(\kappa)$ in  Fig.~\ref{Fig:Mean} for $\kappa d \gg 1$. This reasoning confirms that
the number of contributing modes is the main mechanism for increasing the heat flux at small distances and large filling factors ($f > 0.3$).  

\begin{figure}[Hhbt]
  \epsfig{file=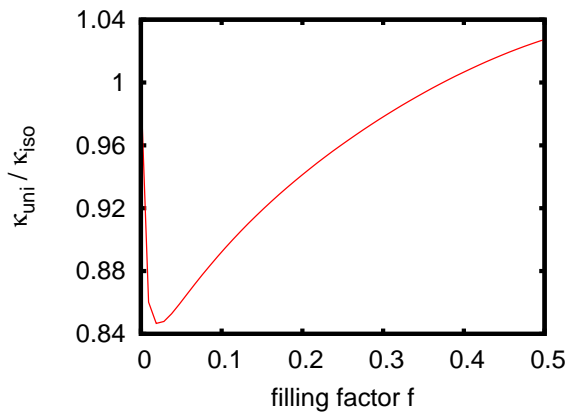, width=0.45\textwidth}
  \caption{\label{Fig:Cutoff} 
           Plot of the normalized cutoff value $\kappa_{\rm uni} / \kappa_{\rm iso}$ over filling factor $f$.}
\end{figure}


%
%

\section{Conclusion}

We have presented a detailed study of near and far field heat transfer
between two flat uniaxial media made of polar materials (in our case, SiC) 
in which cylindrical inclusions drilled orthogonally to surfaces are uniformally
distributed.

After applying the classical stochastic electrodynamic theory to 
anisotropic materials we have shown that, for short distances, the heat flux 
between such media can be significantly larger than those 
traditionally measured between two isotropic materials in 
the same non-equilibrium thermal conditions. For small filling factors
we have determined that this enhancement stems 
from additional surface waves arising at the uniaxial material-vacuum interface, 
clearly indicating that such increase is intrinsically connected to anisotropy. 
Indeed, we did calculations for isotropically rarified SiC plates with low
filling factors ($f\leq0.1$) and found that the
heat transfer modification for is much smaller. In contrast,
for larger filling factors ($f > 0.3$) we have shown that, after
a thorough analysis of the transmission factor, the
enhancement in heat transfer arises mainly from the increased
number of modes contributing to the flux.


%
%

\begin{acknowledgments}
We thank Henri Benisty for very helpful discussions on the subject of homogenization.
S.-A.\ B.\ gratefully acknowledges support from the Deutsche Akademie der Naturforscher Leopoldina
(Grant No.\ LPDS 2009-7). This research was partially supported by Triangle de la Physique, under
the contract 2010-037T-EIEM.
\end{acknowledgments}

%
%

\appendix

\section{Green's dyadic in the gap region}
\label{App:Greensfunction}

In order to construct the Green's dyadic in the vacuum gap we first start
with the Green's dyadic in free space. If $z > z'$ Weyl's expansion for the
Green's dyadic is~\cite{Sipe1986}
\begin{equation}
  \mathds{G}(\mathbf{r,r'}) = \int \!\! \frac{\rd^2 \kappa}{(2 \pi)^2} \, \frac{\ri \re^{\ri \boldsymbol{\kappa}\cdot(\mathbf{x - x'})}}{2 \gamma_\rr} \re^{\ri \gamma_\rr (z - z')} \mathds{1}
\end{equation}
with $\gamma_\rr = \sqrt{\omega^2/c^2 - \kappa^2}$, $\mathbf{x} = (x,y)^{\rm t}$ and $\boldsymbol{\kappa} = (k_x, k_y)^{\rm t}$. The
unit dyadic $\mathds{1}$ is the unit dyadic in the polarization basis and is defined as
\begin{equation}
  \mathds{1} = \hat{\mathbf{a}}^+_\rs \otimes \hat{\mathbf{a}}^+_\rs + \hat{\mathbf{a}}^+_\rp \otimes \hat{\mathbf{a}}^+_\rp.
\end{equation} 
The polarisation vectors for s- and p-polarized waves are given by
\begin{equation}
  \hat{\mathbf{a}}^{+}_\rs = \frac{1}{\kappa} \begin{pmatrix} -k_y \\ k_x \\ 0\end{pmatrix} \qquad \text{and} \qquad
  \hat{\mathbf{a}}^{+}_\rp = \frac{c}{\kappa \omega} \begin{pmatrix} k_x \gamma_\rr \\ k_y \gamma_\rr \\ - \kappa^2\end{pmatrix}.
\end{equation}
By construction both polarization vectors are orthogonal. For propagating waves they are also normalized.
The Fourier component $\mathds{G}(\boldsymbol{\kappa};z,z')$ of the Greens dyadic is defined by
\begin{equation}
  \mathds{G}(\mathbf{r,r'}) = \int \!\!\frac{\rd^2 \kappa}{(2 \pi)^2} \,\mathds{G}(\boldsymbol{\kappa};z,z') \, \re^{\ri \boldsymbol{\kappa}\cdot(\mathbf{x - x'})}. 
\end{equation}

The above expression for the Green's dyadic represents the field of a right going wave at $z$ of a source of unit 
strength placed at $z'$. If a semi-infinite medium is located at $z > d$ then this wave will be reflected so that the
Green's dyadic $\mathds{G}(\boldsymbol{\kappa};z,z')$ reads at $z > z'$
\begin{equation}
   \mathds{G}_A(\boldsymbol{\kappa};z,z') = \frac{\ri}{2 \gamma_\rr} \bigl[ \mathds{1} \re^{\ri \gamma_\rr(z - z')} + \re^{2 \ri \gamma_\rr d} \re^{ \ri \gamma_\rr (z + z')} \mathds{R}_2 \bigr]
\end{equation}
where we have introduced the reflection matrix
\begin{equation}
  \mathds{R}_2 = \sum_{i,j = \{\rs,\rp\} } r_{i,j}^2 \hat{\mathbf{a}}^-_i \otimes \hat{\mathbf{a}}^+_j 
\end{equation}
with the reflection coefficients $r_{i,j}^2$ and the polarization vectors $\hat{\mathbf{a}}^-_p = -c/(\kappa \omega) ( k_x \gamma_\rr, k_y \gamma_\rr, \kappa^2)^{\rm t}$
and $\hat{\mathbf{a}}^-_s = \hat{\mathbf{a}}^+_s$. If there is now a second semi-infinite medium at $z < 0$ with a reflection operator defined as
\begin{equation}
  \mathds{R}_1 = \sum_{i,j = \{\rs,\rp\}} r_{i,j}^1 \hat{\mathbf{a}}^+_i \otimes \hat{\mathbf{a}}^-_j 
\end{equation}
the waves in that cavity will be multiply reflected at the boundaries at $z = 0$ and $z = d$ so that \cite{Philbin08}
\begin{equation}
\begin{split}
  \mathds{G}_A(\boldsymbol{\kappa};z,z') &= \frac{\ri}{2 \gamma_\rr} \biggl[ \mathds{1} \re^{\ri \gamma_\rr(z - z')} + \re^{2 \ri \gamma_\rr d} \re^{ -\ri \gamma_\rr (z + z')} \mathds{R}_2  \\
                                         &\qquad + \re^{2 \ri \gamma_\rr d} \re^{ \ri \gamma_\rr (z - z')} \mathds{R}_1 \mathds{R}_2 \\
                                         &\qquad +  \re^{4 \ri \gamma_\rr d} \re^{ -\ri \gamma_\rr (z + z')} \mathds{R}_2 \mathds{R}_1 \mathds{R}_2 
                                         + \ldots \biggr]. 
\end{split}
\end{equation}
Summing up all contributions we get
\begin{equation}
\begin{split}
   \mathds{G}_A(\boldsymbol{\kappa};z,z')&= \frac{\ri}{2 \gamma_\rr} \biggl[ 
                                            \mathds{D}^{12} \re^{\ri \gamma_\rr (z - z')} \\
                                         &\qquad + \mathds{D}^{21} \mathds{R}_2 \re^{2 \ri \gamma_\rr d} \re^{-\ri \gamma_\rr (z + z')} \biggr]
\end{split}
\label{Eq:GreenA}
\end{equation}
where we have introduced
\begin{align}
    \mathds{D}^{12} = (\mathds{1} - \mathds{R}_1 \mathds{R}_2 \re^{2 \ri \gamma_\rr d})^{-1}, \\
    \mathds{D}^{21} = (\mathds{1} - \mathds{R}_2 \mathds{R}_1 \re^{2 \ri \gamma_\rr d})^{-1}.
\end{align}

The expression in Eq.~(\ref{Eq:GreenA}) is not yet the complete intracavity Green's dyadic, since we have not considered the
waves which start from $z'$ as left going waves and arrive after being reflected at the boundary at $z = 0$ at $z > z'$.
With the same reasoning as for $ \mathds{G}_A(\boldsymbol{\kappa};z,z')$ we find for this contribution
\begin{equation}
  \begin{split}
   \mathds{G}_B(\boldsymbol{\kappa};z,z')&= \frac{\ri}{2 \gamma_\rr} \biggl[ 
                                            \mathds{D}^{12} \mathds{R}_1 \re^{\ri \gamma_\rr (z + z')} \\
                                         &\qquad + \mathds{D}^{21} \mathds{R}_2 \mathds{R}_1 \re^{2 \ri \gamma_\rr d} \re^{\ri \gamma_\rr (z' - z)} \biggr]
\end{split}
\label{Eq:GreenB}
\end{equation} 
Finally, the intracavity Green's dyadic is given by the sum of Eq.~(\ref{Eq:GreenA}) and (\ref{Eq:GreenB}) yielding
\begin{equation}
\begin{split}
  \mathds{G}_{\rm intra} &= \frac{\ri}{2 \gamma_\rr} \biggl[ \mathds{D}^{12} \biggl( \mathds{1} \re^{\ri \gamma_\rr (z - z')} + \mathds{R}_1 \re^{\ri \gamma_\rr (z + z')} \biggr) \\
                         &\qquad\quad +  \mathds{D}^{21} \biggl( \mathds{R}_2 \mathds{R}_1 \re^{\ri \gamma_\rr (z' - z)} \re^{2 \ri \gamma_\rr d} \\ 
                         &\qquad\qquad\qquad + \mathds{R}_2 \re^{2 \ri \gamma_\rr d} \re^{-\ri \gamma_\rr (z + z')} \biggr) \biggr]
\end{split}
\label{Eq:IntracavityGreen}
\end{equation}

\section{The resonances in an anisotropic material}

The precise location of resonances can be analytically determined from expression (\ref{Eq:Disp}) by solving
\begin{equation}
  \epsilon_\parallel \epsilon_\perp = 1.
\end{equation}
Frequencies which satisfy this condition are resonance frequencies of medium because they correspond to a flat region of the dispersion curve in the $(\omega,\kappa)$ plane and therefore to strong density of states. Using expressions (\ref{Eq:eps_par}) and (\ref{Eq:eps_perp}) this equation can be 
recast into the following form 
\begin{equation}
 a \epsilon_\rh^3 + b \epsilon_\rh^2 + c \epsilon_\rh + d = 0,
\end{equation}
with
\begin{align}
  a &= (1 - f)^2, \\
  b &= \epsilon_\ri (1 - f) (2 f + 1), \\
  c &= (1 + f) (\epsilon_\ri^2 f - 1), \\
  d &= \epsilon_\ri (f - 1).
\end{align}
The solutions of this equation can readily been calculated using the Cardano's method~\cite{Cooke2008}. When inclusions are made by pure vacuum (i.e. $\epsilon_\ri = 1$), these solutions are real and read
\begin{equation}
  \epsilon_{h,n} = 2 \sqrt{\frac{-p}{3}} \cos\biggl[ \frac{1}{3} \arccos\biggl( -\frac{q}{2}\sqrt{\frac{27}{-p^3}} \biggr) + \frac{2 \pi n}{3} \biggr]
\end{equation}
for $n = 0,1,2$.
with
\begin{equation}
  p = \frac{3 a c - b^2}{3 a^2} \qquad\text{and} \qquad q = \frac{27 a^2 d - 9 a b c + 2 b^3}{27 a^3}
\end{equation}
Only one of these solutions is positive and must be conserved to search the resonance frequencies.

%
%

\end{document}